\documentclass[prl,twocolumn,groupedaddress,amssymb]{revtex4}
\usepackage{graphicx}
\usepackage{bm}

\begin{document}
\title{Dissipation in the superconducting state of $\kappa$-(BEDT-TTF)$_2$Cu(NCS)$_2$}
\author{Liang Yin}
\author{Moon-Sun Nam}
\author{James G.\ Analytis}
\author{Stephen J.\ Blundell}
\author{Arzhang Ardavan}
\affiliation{Department of Physics, Oxford University, Oxford, UK}
\author{John A.\ Schlueter}
\affiliation{Materials Science Division, Argonne National Laboratory, Argonne, IL 
60439, USA}
\author{Takahiko Sasaki}
\affiliation{Institute for Materials Research, Tohoku University, Sendai 980-8577, Japan}

\date{\today}

\begin{abstract}
We have studied the interlayer resistivity of the prototypical quasi-two-dimensional organic superconductor $\kappa$-(BEDT-TTF)$_2$Cu(NCS)$_2$ as a function of temperature, current and magnetic field, within the superconducting state. We find a region of non-zero resistivity whose properties are strongly dependent on magnetic field and current density. There is a crossover to non-Ohmic conduction below a temperature that coincides with the 2D vortex solid -- vortex liquid transition. We interpret the behaviour in terms of a model of current- and thermally-driven phase slips caused by the diffusive motion of the pancake vortices which are weakly-coupled in adjacent layers, giving rise to a finite interlayer resistance.
\end{abstract}
\maketitle

One of the many interesting observations made in layered superconductors is the pronounced broadening of the superconducting transition in a magnetic field, and dissipation within the superconducting state~\cite{zettl:1991,mckay:1990}. These effects have been studied extensively in high-T$_{\mathrm c}$ superconductors, and various mechanisms have been discussed to explain them, including vortex motion~\cite{Palstra:1990}, vortex-antivortex excitations~\cite{martin:1989,stamp-comment:1989}, and Josephson coupling between layers~\cite{zettl:1991}. In general, dissipation of intralayer currents arises from vortex motion, while dissipation of interlayer currents has its origins in phase fluctuations between neighbouring Josephson-coupled layers.

Despite the interest paid to these effects in high-$T_{\mathrm c}$ superconductors, there are few studies of dissipation within the superconducting state in organic superconductors. A very thorough study of dissipation in the intralayer conductivity within the superconducting state of the prototype quasi-two-dimensional organic superconductor $\kappa$-(BEDT-TTF)$_2$Cu(NCS)$_2$~\cite{saito:1998} was performed by Sasaki and coworkers~\cite{sasaki:2002}; they identified a region of non-linear intralayer conductivity magnetic fields close to the critical field and at temperatures low compared to the critical temperature, which was interpreted as evidence for a novel vortex ``slush'' state characterised by quantum fluctuations of the vortex lattice. Others have investigated the resistive transition in the presence of a magnetic field, and found that it is consistent with vortex liquid behaviour~\cite{Friemel:1996,Friemel:1997}. 

In this paper, we examine in detail the dissipation of interlayer currents within the superconducting state in $\kappa$-(BEDT-TTF)$_2$Cu(NCS)$_2$. In our experiment the magnetic field and current are parallel, and aligned perpendicular to the crystal planes; in this configuration, there is no Lorentz force on the vortices. We find a region within the superconducting state which exhibits a non-zero resistivity that is strongly dependent on the magnetic field and current density. We observe non-ohmic conduction below a temperature $T^*$, which is close to the vortex melting transition temperature. We correlate these effects with the temperature at which global intralayer phase coherence is established, as measured by the Meissner effect in the same sample.

High quality single crystal samples, grown by electrochemical techniques~\cite{saito:1998}, were obtained from several sources. The samples were typically of dimensions 1~mm$\times$0.5~mm$\times$0.2~mm; the smallest dimension reliably corresponds to the interlayer direction. Resistance was measured using the standard four-probe technique. Two contacts were made to each of the large surfaces of the crystal using gold wire and graphite paste. In order to ensure good thermal contact between the sample and the thermometer, both were mounted on a single crystal of quartz (which is a good thermal conductor at low temperatures) with thermally conductive grease. An a.c.\ current of between 1 and 50~$\mu$A (corresponding to current densities of the order of 1 to 50 Am$^{-2}$) was applied at 77.7~Hz; the voltage was detected using an EG\&G lockin amplifier. Measurements were performed in a $^4$He flow cryostat, equipped with a 17~T Nb$_{3}$Sn superconducting solenoid. Zero magnetic field measurements were performed following a thermal cycling of the solenoid above its $T_{\mathrm c}$ = 18~K, to release remnant flux.

The sample was cooled from room temperature to 20K at a rate less than 1~K/min~\cite{su:1998}. The interlayer resistance was then measured as a function of magnetic field and current, while the sample was slowly cooled to 1.8K at a rate of $\sim$0.5~K/min.

\begin{figure}
\includegraphics[width=8cm]{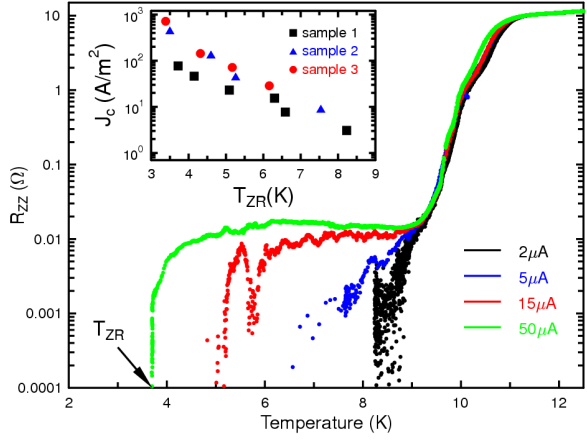}
\caption{(Colour online) Temperature dependence of interlayer resistance in zero applied magnetic field, for a range of currents. Inset: critical current density for the onset of dissipation, $J_c$, as a function of T$_{ZR}$, the temperature at which the resistance becomes zero. 
}
\label{fig:tempsweep}
\end{figure}

\begin{figure}
\includegraphics[width=8cm]{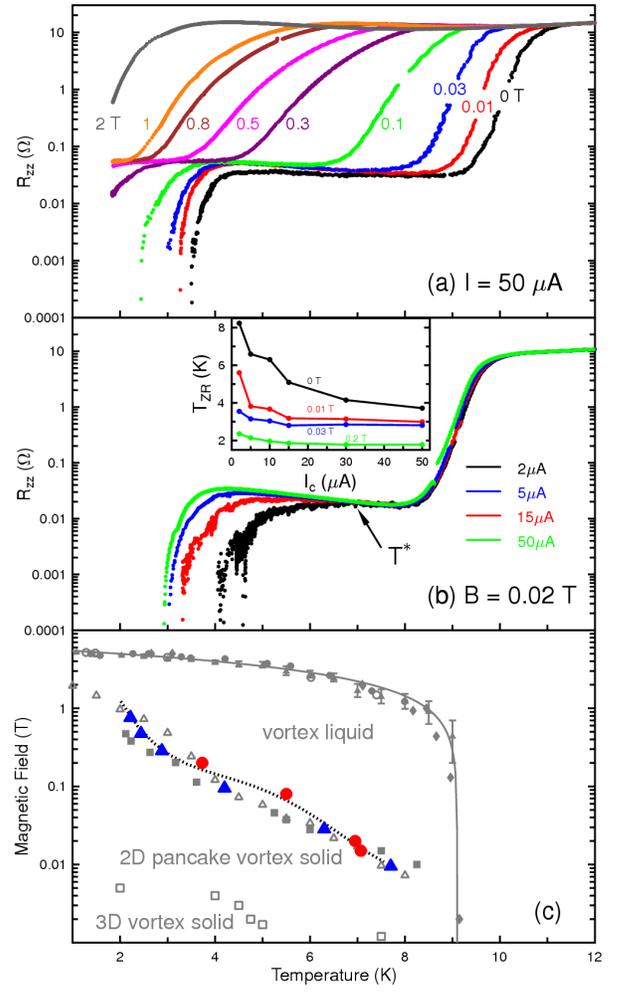}
\caption{(Colour online) (a) The temperature dependence of the interlayer resistance in a range of magnetic fields $B$, applied perpendicular to the conducting planes. Traces are shown for $B$ = 0~T (highest temperature transition), 0.01~T, 0.03~T, 0.1~T, 0.3~T, 0.5~T, 0.8~T, 1~T and 2~T. The current is 50~$\mu$A. The data are from sample 2.
(b) The temperature dependence of the interlayer resistance in a magnetic field $B$ = 0.02~T, for a range of currents $I$. Traces are shown for $I$ = 2~$\mu$A, 5~$\mu$A, 15~$\mu$A and 50~$\mu$A. The I-V response is linear (i.e.\ the conductivity is Ohmic) above a characteristic temperature T$^{*}$. Inset to (b): $T_{ZR}$ as a function of $I_c$ for a range of magnetic fields. (c) In grey, the phase diagram of $\kappa$-(BEDT-TTF)$_2$Cu(NCS)$_2$ adapted from reference~\cite{singleton:2002}; see text for explanation of data points. $T^*$ is shown for a range of magnetic fields for sample 2 (blue triangles) and sample 3 (red circles); $T^*$ is coincident with the 2D vortex lattice melting line.
}
\label{fig:tempsweepB}
\end{figure}

Figure~\ref{fig:tempsweep} shows the temperature dependence of the interlayer resistance in zero magnetic field, with applied currents of 2, 5, 15 and 50$\mu$A (RMS) at 77.7Hz. The transition occuring between 9~K and 11~K is the superconducting transition, as reported by others~\cite{ishiguro-yamagi}. At temperatures below 9~K, there exists a state with low but non-zero resistance. This low-resistance state is strongly dependent on current for temperatures between 4~K and 9~K. For small currents, a zero resistance state is established at temperatures very close to the superconducting transition. However, as the current increases the temperature at which zero resistance occurs, $T_{ZR}$, becomes lower, falling to below 4~K for a current of 50~$\mu$A. Furthermore, the resistance is apparently not monotonic with temperature in this region; there is a range of temperatures for which ${\mathrm d}R/{\mathrm d}T$ is negative. Note that the superconducting transition between 9~K and 11~K is to a large extent independent of the applied current, allowing us to rule out Joule heating as the cause of the current-dependent finite resistance state at lower temperatures. The inset to figure \ref{fig:tempsweep} shows the critical current density, $J_c$, (at which zero resistance occurs) as a function of the temperature $T_{ZR}$. While $J_c$ exhibits some sample dependence, a general trend is clear: in all samples it increases approximately exponentially as $T$ falls.

A zero magnetic field current-dependent resistive state can be explained in terms of a thermally-activated Josephson effect, proposed by Ambegaokar and Halperin~\cite{ambegaokar:1969} in 1969. They consider the effect of thermal fluctuations in the vicinity of the superconducting transition, which can lead to phase decoherence between the Josephson-coupled superconducting layers, and thus to interlayer dissipation. An interlayer current serves to decouple the layers further, so that the thermal-fluctuation-induced decoupling persists to lower temperatures. This model may be appropriate for $\kappa$-(BEDT-TTF)$_2$Cu(NCS)$_2$, with its highly two-dimensional structure, large anisotropy parameter and short interlayer coherence length~\cite{saito:1998}.

Given the large anisotropy parameter, with this experiment we cannot rule out the possibility that the effect is associated with Kosterlitz-Thouless vortex-antivortex pairs~\cite{kosterlitz:1973,martin:1989,stamp-comment:1989}; dissipation arises from the pair-breaking effect of intralayer components of the current. Confirmation of this scenario would require measurement of the temperature dependence of the intraplane conductivity. Nevertheless, others have invoked Kosterlitz-Thouless-like scenarios to explain their observations in other quasi-two-dimensional organic superconductors such as $\kappa$-(BEDT-TTF)$_4$Hg$_{2.89}$Br$_{8}$ \cite{Kinoshita:1999} and $\alpha$-(BEDT-TTF)$_{2}$NH$_{4}$Hg(SCN)$_{4}$ \cite{sato:1995}.

However, it should be noted that neither of these models has been shown to give rise to a negative ${\mathrm d}R/{\mathrm d}T$ as we observe.

Figure~\ref{fig:tempsweepB}(a) shows the temperature dependence of the interlayer resistance in the presence of a magnetic field $B$ applied perpendicular to the layers, for a range of fields 0.01~T~$< B <$~2~T; the current is 50~$\mu$A. The magnetic field stabilises the non-zero-resistance state, with $T_{ZR}$ decreasing as the field is applied. For a field of 0.5~T, $T_{ZR}$ falls below the lowest temperature studied here, 1.8~K. For small fields, the region of negative ${\mathrm d}R/{\mathrm d}T$ is extended over the zero-field case. For larger fields the superconducting transition moves down in temperature rapidly and the non-zero-resistance state contracts.

Figure~\ref{fig:tempsweepB}(b) shows the temperature dependence of the interlayer conductivity in the presence of a magnetic field of 0.02~T for a range of currents 2~$\mu$A~$< I <$~50~$\mu$A. The notable feature of these data is that the traces are coincident at high temperatures, and diverge (indicating the onset of non-Ohmic conduction) below a temperature $T^*$ ($\approx$7~K for $B = $0.02~T as shown in figure~\ref{fig:tempsweepB}(b)), somewhat below the superconducting transition temperature. 

In figure~\ref{fig:tempsweepB}(c) we compare $T^*$ measured at a range of magnetic fields (blue triangles: sample 2, red circles: sample 3) with the vortex phase diagram obtained from other experiments on the same material (data points in grey, adapted from~\cite{singleton:2002}) . The $H_{c2}$ data were obtained from MHz penetration studies (filled triangles)~\cite{singleton:2002}, microwave penetration studies (filled circles)~\cite{js155}, thermal conductivity (open circles)~\cite{js155} and magnetisation (filled diamonds)~\cite{sasaki:1998,js86}. The vortex liquid -- 2D vortex solid transision is found from the irreversibiliity field in magnetisation (open triangles) and studies of vortex melting in magnetometry and microwave response~\cite{mola:2001,js96}. The 3D vortex melting line was obtained from muon-spin rotation studies~\cite{js104}. We find that $T^*$ coincides with the 2D vortex solid -- vortex liquid melting line, suggesting a simple explanation for the qualitative behaviour that we find.

In the vortex liquid state, pancake vortices are fully mobile within the planes. Thus the degree of correlation (or lack thereof) between the phases in neighbouring planes is unaffected by a current, and the finite conductivity arising from the phase slips between planes is Ohmic. (This scenario is related to a model developed by Koshelev~\cite{koshelev:1995} that describes the in-plane dissipation arising from the thermally activated diffusive motion of pancake vortices.) However, once the temperature falls below the point where a 2D vortex lattice establishes itself in each plane, the weak interlayer correlations between these lattices become sensitive to the interlayer current, causing a finite but non-Ohmic interlayer conductivity. The fact that ${\mathrm d}R/{\mathrm d}T$ is rather strongly negative for large currents below $T^*$ suggests that the in-plane ordering of vortices allows larger phase-slips between planes to occur than for the vortex liquid case. The origin of this effect is not yet clear, but one might, for example, speculate that it is possible for the positions of pancake vortices in two neighbouring 2D lattices to become {\em anti}correlated, rather than just {\em un}correlated in the vortex liquid case. However, a quantitive description of this effect is beyond the scope of this paper. 

The inset to figure~\ref{fig:tempsweepB}(b) shows $T_{ZR}$ as a function of $I_C$ for a range of magnetic fields. $T_{ZR}$ can be suppressed monotonically by either a magnetic field or a current. However, as the magnetic field is increased, the effect of the current is diminished. In the qualitative picture described above, this can be interpreted as indicating that both magnetic field and current serve to induce decoherence of the superfluids in adjacent layers.

\begin{figure}
\includegraphics[width=8cm]{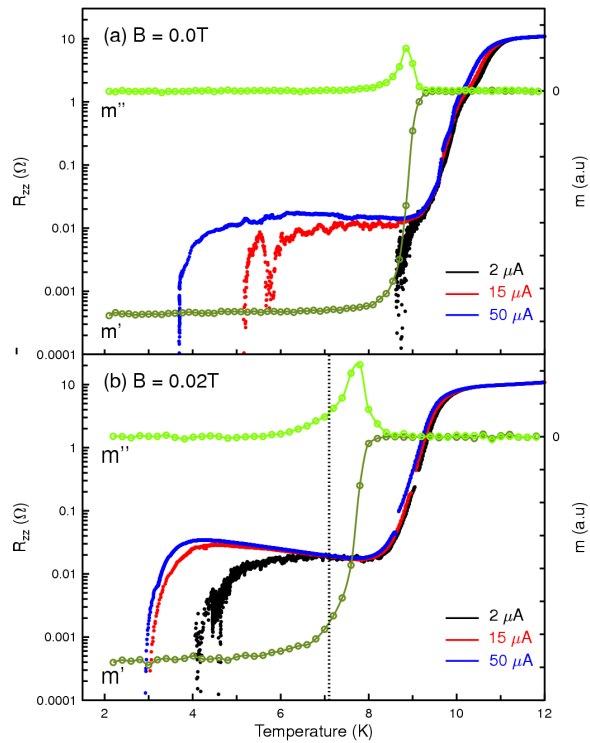}
\caption{(Colour online) Real (dark green) and imaginary (light green) parts of the magnetization (right-hand scale), extracted from a.c.\ susceptibility measurements in (a) zero static magnetic field and (b) 0.02~T applied perpendicular to the conducting planes. Also shown are measurements of the interlayer conductivity (left-hand scale) on the same sample in similar magnetic field conditions for a range of currents. The dotted vertical line in (b) marks $T^*$.}
\label{fig:BD}
\end{figure}

Further insight may be gained by comparing $T^*$ with the temperature at which the Meissner effect is established. Figure~\ref{fig:BD}(a) shows the real and imaginary parts of the magnetisation, obtained from an a.c.\ susceptibility measurement (using a Quantum Design SQUID magnetometer with an a.c.\ field strength of 30~$\mu$T at a frequency 990~Hz), as a function of temperature in zero static magnetic field. Also shown is the resistivity measured for a range of currents. The Meissner effect is essentially complete at the onset of the zero-resistance state as measured with a small current. Figure~\ref{fig:BD}(b) shows the same measurement in a static field of 0.02~T applied perpendicular to the sample planes. In contrast to the zero-field case, the Meissner effect is apparently complete at temperatures higher than the onset of zero resistance, even for small applied currents. Indeed, the saturation of the magnetization is well correlated with $T^*$; this indicates that, at least for static fields applied perpendicular to the superconducting planes, the Meissner effect is established once the intralayer 2D vortex lattice is stabilised, and is not dependent on interlayer phase coherence. This supports the use of the Meissner effect as a probe of vortex lattice melting and our assignment of $T^*$, derived from resistivity measurements, as the temperature at which this occurs. 

In previous work, the irreversibility line has been studied using magnetization probes~\cite{mola:2001,sasaki:1998}. While there has been some discussion of the relationship between the irreversibility line and the 2D vortex lattice melting line~\cite{sasaki:1998}, and there is some evidence for their deviation at very low temperatures (the quantum regime)~\cite{mola:2001}, at higher temperatures (in the classical regime studied here) they are thought to be approximately coincident~\cite{singleton:2002}.

In conclusion, we have examined dissipation of interlayer currents within the superconducting state of the highly layered superconductor $\kappa$-(BEDT-TTF)$_2$Cu(NCS)$_2$ in the temperature range down to about $0.2 T_c$ and as a function of magnetic field. We interpret the existence of a low (but non-zero) resistance state as arising from phase decoherence between the superfluids in adjacent layers. Above a characteristic temperature $T^{*}$, which is magnetic field dependent and apparently coincident with the 2D vortex lattice melting line, the resistivity is Ohmic. Below $T^*$, a non-Ohmic region is interpreted as evidence that the coherence between vortex lattices in neighbouring layers is current-dependent.

This work is supported by EPSRC. AA is supported by the Royal Society. The authors would like to thank J.~Singleton for providing data presented in figure~\ref{fig:tempsweepB}(c) and helpful discussions. Work at Argonne National Laboratory was supported by the Office of Basic Energy Sciences, Division of Materials Science, U.S. Department of Energy under contract no. DE-AC02-06CH11357.

\end{document}